\documentclass[a4paper,10pt,twoside]{cpc-hepnp}

\usepackage{multicol}
\usepackage{graphicx}
\usepackage{booktabs}
\usepackage{amssymb,bm,mathrsfs,bbm,amscd}
\usepackage[tbtags]{amsmath}
\usepackage{lastpage}
\usepackage{CJK}
\usepackage{subfigure}

\begin{document}
\begin{CJK*}{GBK}{song}

\fancyhead[co]{\footnotesize CHEN Si~ et al: Beam breakup simulation study for high energy ERL}

\footnotetext[0]{Received 15 February 2014}

\title{Beam breakup simulation study for high energy ERL\thanks{Supported by the Major State Basic Research Development Program of China under Grant No. 2011CB808404.}}

\author{
      CHEN Si $^{1,2}$
      \quad M. Shimada$^{3}$
      \quad N. Nakamura$^{3}$\\
      \quad HUANG Sen-Lin$^{2;1)}$\email{huangsl@pku.edu.cn}
      \quad LIU Ke-Xin$^{2;1)}$\email{kxliu@pku.edu.cn}
      \quad CHEN Jia-Er$^{1,2}$
}
\maketitle

\address{%
$^1$ School of Physics, University of Chinese Academy of Sciences, Beijing 100190, China\\
$^2$ State Key Laboratory of Nuclear Physics and Technology,
Institute of Heavy Ion Physics, \\Peking University, Beijing 100871, China\\
$^3$ KEK, Oho 1-1 Tsukuba, Ibaraki 305-0801, Japan\\
}

\begin{abstract}
The maximum beam current can be accelerated in an Energy Recovery Linac (ERL) can be severely limited by the transverse multi-pass beam breakup instability (BBU), especially in future ERL light sources with multi-GeV high energy beam energy and more than 100 mA average current. In this paper, the multi-pass BBU of such a high energy ERL is studied based on the simulation on a 3-GeV ERL light source proposed by KEK. It is expected to provide a reference to the future high energy ERL projects by this work.

\end{abstract}



\begin{multicols}{2}

\section{Introduction}
Energy recovery linacs (ERLs) are expected to provide high current electron beams with an RF power supply much lower than traditional linacs. At the same time, the excellent beam qualities of linear accelerator, e.g., low emittance, small energy spread and short bunch length, are able to be maintained compared with those of storage rings. These characteristics make ERLs very suitable for future ultra-brightening synchrotron light sources, free electron lasers, nuclear physics researches, and so on.

Multi-pass transverse beam breakup is known as one of the key issues of ERLs. It is primarily contributed by a positive feedback between the recirculated bunch with transverse offset and insufficiently damped dipole higher order modes in superconducting cavity. The average current of ERL can be severely limited by this effect. Studies on the multi-pass BBU of small-scale ERLs with several tens MeV and average current around 10 mA have been done before\cite{lab1,lab2}. For ERL based synchrotron light sources with the energy of a few GeV, hundreds of cavities will be used and usually an average current of more than 100 mA is required. In that case, multi-pass BBU is a more significant issue and should be analyzed carefully.

Several high energy ERL light sources are proposed in the world\cite{lab3}. One of them is a synchrotron X-ray light source based on a 3-GeV ERL at KEK, which is expect to be a successor of the existing synchrotron light sources of Photon Factory in KEK. A preliminary design report of this project has been published in 2012 \cite{lab4,lab5}. Recently, we performed the study of multi-pass beam breakup for this facility. In this paper, the BBU simulation results of the KEK 3-GeV ERL are firstly presented. Some features of the BBU of high energy ERLs are then discussed based on the simulation results.



\section{Multi-pass beam breakup}

In ERLs, an electron bunch deflected by an dipole HOM on the first pass comes back to the same cavity again on the second or higher passes with a transverse offset. The recirculating bunch can constructively or destructively interact with the HOM which deflected it on the previous pass. Therefore, there exists a feedback between the HOM field and the recirculating bunch. The enhancing feedback by a series of bunches can cause an exponential increase of the HOM power if the HOM is not sufficiently damped. The HOM transverse kick will become strong enough so that the beam strikes the cavity wall and gets lost. This process is called multi-pass beam breakup.

A two-dimensional analytical formula for the multi-pass BBU threshold current is \cite{lab6}
\begin{eqnarray}
I_{th}=-\frac{2pc}{e(\frac{\omega}{c})(\frac{R_d}{Q})Q_{ext} M_{12}^*\sin(\omega T_r)},
\label{eq1}
\end{eqnarray}
where $(R_d/Q)$ is the shunt impedance of the dipole mode in the cavity, $Q_{ext}$ is the external quality factor, $\omega$ is the HOM frequency, $T_r$ is the bunch recirculating time, and
$$M_{12}^*=T_{12}\cos^2{\theta}+\frac{1}{2}(T_{14}+T_{23})\sin{2\theta}+T_{34}\sin^2{\theta},$$
where $T_{ij}$ are the elements of the pass-to-pass transport matrix and $\theta$ is the polarization angle of the dipole HOM.

Eq.~\ref{eq1} shows the main determinants of multi-pass BBU instability in an ERL. But it's only valid in the case of single cavity, single HOM and $M_{12}^*\sin(\omega T_r)<0$. In real cases, the situation is more complicated. It's necessary to use simulation codes to compute the BBU threshold current. In this paper, the particle tracking code $bi$ developed by I. Bazarov\cite{lab7} at Cornell University is used in the simulation.

\section{BBU simulation}
\subsection{KEK 3-GeV ERL light source}

\begin{center}
\includegraphics[width=7.0cm]{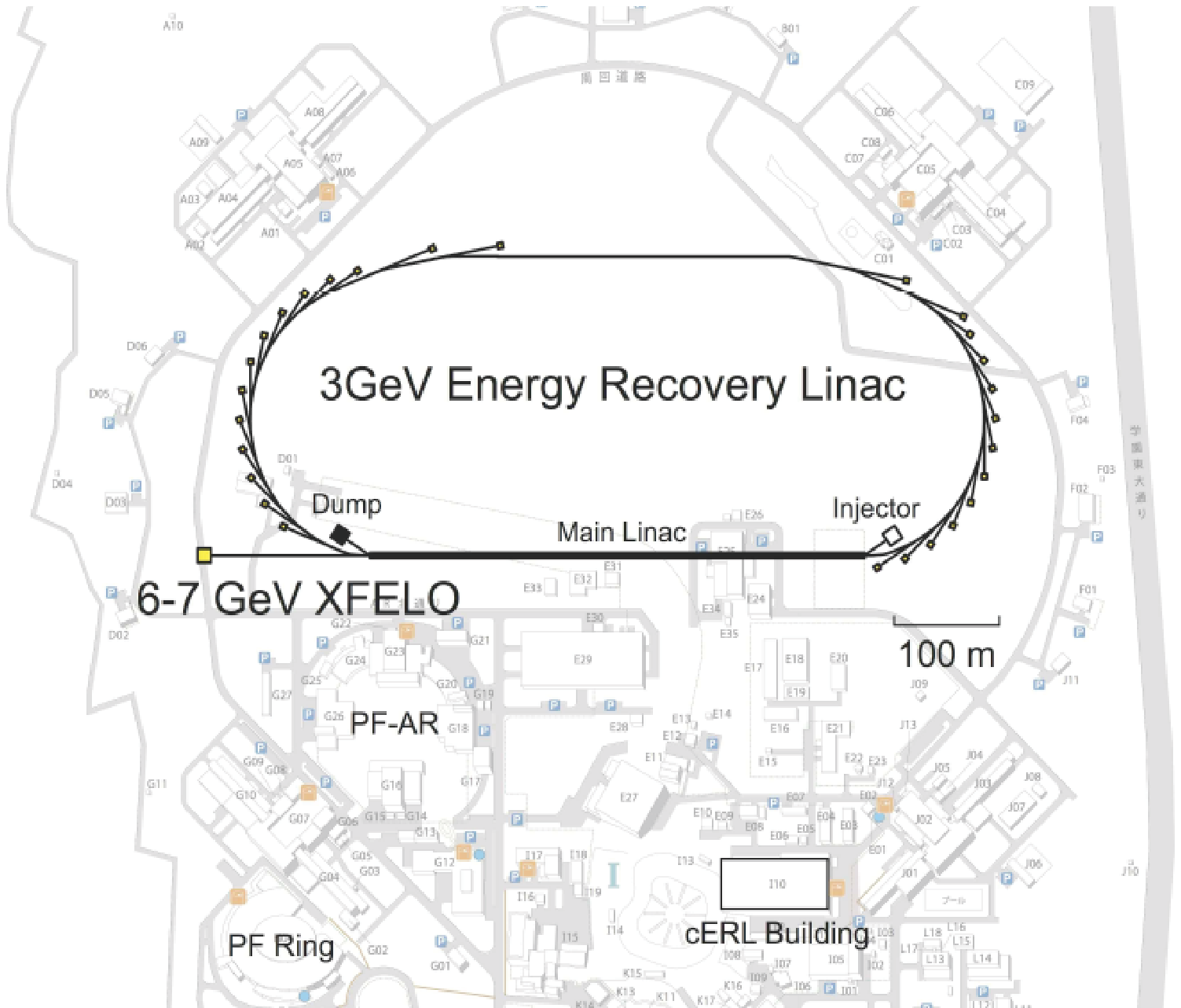}
\figcaption{\label{fig1}   A tentative layout of the 3-GeV ERL light source with an XFEL-O located at KEK Tsukuba campus}
\end{center}

An overall layout of the KEK 3-GeV ERL light source is shown in Fig.~(\ref{fig1}). The electron beam is injected at the energy of 10 MeV and accelerated by the main linac to about 3 GeV. One of the linac configuration consists of 28 cryomodules with 8 cavities in each cryomodule with an accelerating gradient of $13.4$ MV/m and the final energy is 3.01 GeV \cite{lab4}. Another configuration consists of 34$\times$8 cavities with an accelerating gradient of $12.5$ MV/m and the final energy is 3.41 GeV \cite{lab11}. The betatron function and dispersion of the first linac configuration are shown in Fig.~(\ref{fig2}). 


\begin{center}
\includegraphics[width=8.0cm]{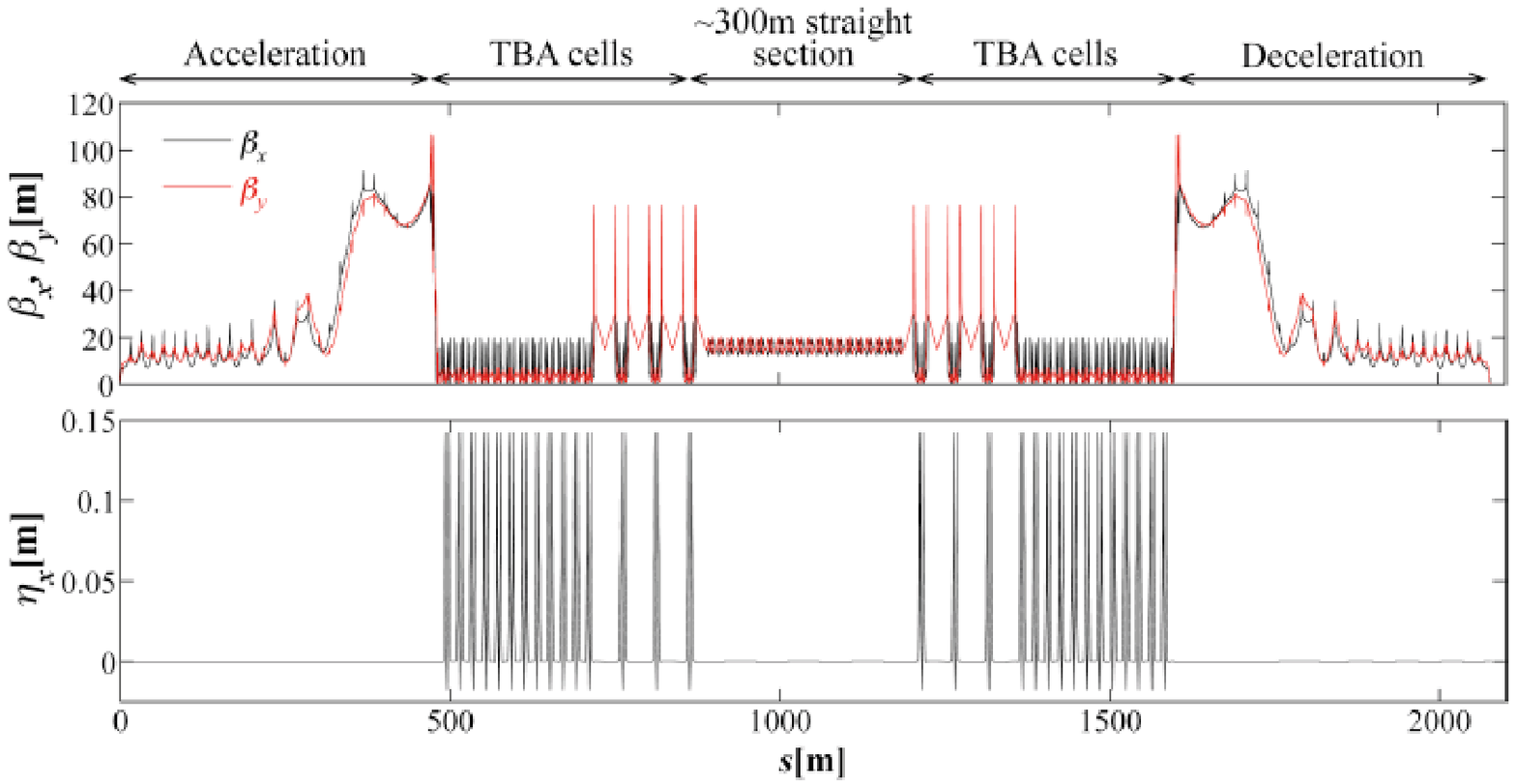}
\figcaption{\label{fig2}   Betatron function (upper) and dispersion (lower) of 3-GeV ERL light source }
\end{center}

To improve the dipole HOM damping, a 9-cell KEK-ERL mode-2 cavity, which has a large iris with the diameter of 80 mm and two large beam pipes with the diameters of 100 and 123 mm, has been developed \cite{lab8}. A previous work shows the BBU threshold current of more than 600 mA can be achieved when applying this type of cavity to a 5-GeV ERL design\cite{lab9}. Several major HOMs in the mode-2 cavity are listed in Table~\ref{tab1}.

\begin{center}
\includegraphics[width=8.0cm]{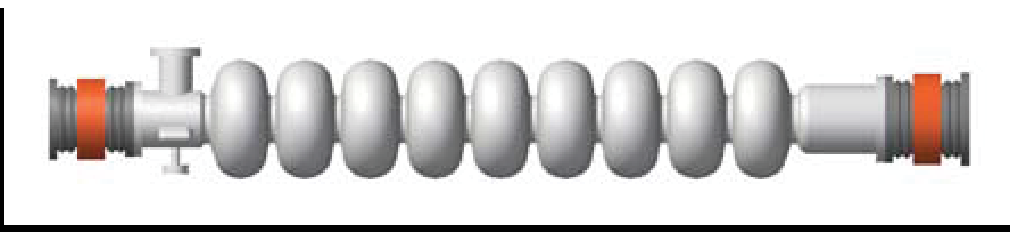}
\figcaption{\label{fig3}   1.3 GHz 9-cell KEK-ERL mode-2 cavity}
\end{center}

\small
\begin{center}
\tabcaption{ \label{tab1}  Major HOMs in KEK-ERL 9-cell cavity}
\begin{tabular*}{75mm}{c@{\extracolsep{\fill}}ccc}
\toprule $f$  &   $Q_e$   &   $R/Q$      &   $(R/Q)Q_e\cdot f$\\
$GHz$ &  &$\Omega/cm^2$ & $\Omega/cm^2/GHz$ \\
\hline
 1.835  &   1.1010$\times10^3$   &   8.087   &   4852    \\
 1.856  &   1.6980$\times10^3$   &   7.312   &   6691    \\
 2.428  &   1.6890$\times10^3$   &   6.801   &   4732    \\
 3.002  &   2.9990$\times10^4$   &   0.325   &   3246    \\
 4.011  &   1.1410$\times10^4$   &   3.210   &   9135    \\
 4.330  &   6.0680$\times10^5$   &   0.018   &   2522    \\

\bottomrule
\end{tabular*}
\end{center}
\normalsize

\subsection{Influence of betatron phase advance to BBU threshold current}
As can be seen in Eq.~\ref{eq1}, the BBU threshold current is a function of $M_{12}^*$. For simplicity, we assume that there is no x-y coupling in the recirculating loop and the HOMs have two different directions of polarizations ($x$ ($\theta=0^\circ$) and $y$ ($\theta=90^\circ$)) but with the same value of frequency, $R/Q$ and $Q_{ext}$. In this case, the value of $M_{12}^*$ is only a function of $T_{12}$ or $T_{34}$ for the two independent polarizations, respectively. The value of $T_{12}$($T_{34}$) between a region with momentum $p_i$ and a region with momentum $p_f$ can be written in terms of $\beta$-function and phase advance $\Delta\psi$ as
\begin{equation}
\label{eq2}
T_{12}(T_{34})(i\rightarrow f)=\sqrt{\frac{\beta_i\beta_f}{p_ip_f}}\sin{\Delta\psi}.
\end{equation}

In order to simulate the transverse dynamics correctly, it is important to include the focusing effect of the RF field in the superconducting cavity. In this work, the Rosenzweig's form of the transport matrix for a pure $\pi$-mode standing-wave cavity \cite{lab10} is applied in the simulation, i.e.,
\small
\begin{eqnarray}
\label{eq5}
M_{cav}=
\left(\begin{matrix}\cos{\alpha}-\sqrt{2}\sin{\alpha}&\sqrt{8}\frac{\gamma_i}{\gamma^{'}}\sin{\alpha}\\
-\frac{3}{\sqrt{8}}\frac{\gamma^{'}}{\gamma_f}\sin{\alpha}&\frac{\gamma_i}{\gamma_f}\left[\cos{{\alpha}}+\sqrt{2}\sin{\alpha}\right]\end{matrix}\right),
\end{eqnarray}
\normalsize
where $\alpha=\frac{1}{\sqrt{8}}\ln{\frac{\gamma_f}{\gamma_i}}$, $\gamma_{i(f)}$ is the initial (final) relativistic factor of the particle, $\gamma^{'}=qE_{0}\cos(\Delta{\phi})/m_{0}c^2$ where $E_0$ is the maximum particle energy gain from the RF cavity and $\Delta\phi$ is the phase of acceleration.



For higher BBU threshold current, one has to make the pass-to-pass value of $T_{12}$ ($T_{34}$)  as smaller as possible. As shown in Eq.~\ref{eq2}, by adjusting the betatron phase advances to make its value to be an integer of multiple of $\pi$ throughout the whole recirculating loop, $T_{12}(T_{34})=0$ can be achieved. An extremely large BBU threshold current up to infinite is obtained in the single cavity case consequently. In real ERL configurations with more than one cavity, the ideal condition of $\Delta \psi=0$ cannot be satisfied for every cavity in the linac. Yet, still we can scan the betatron phase advance of the return loop to search for the optimized value of BBU threshold current.



\begin{center}
\includegraphics[width=6.5cm]{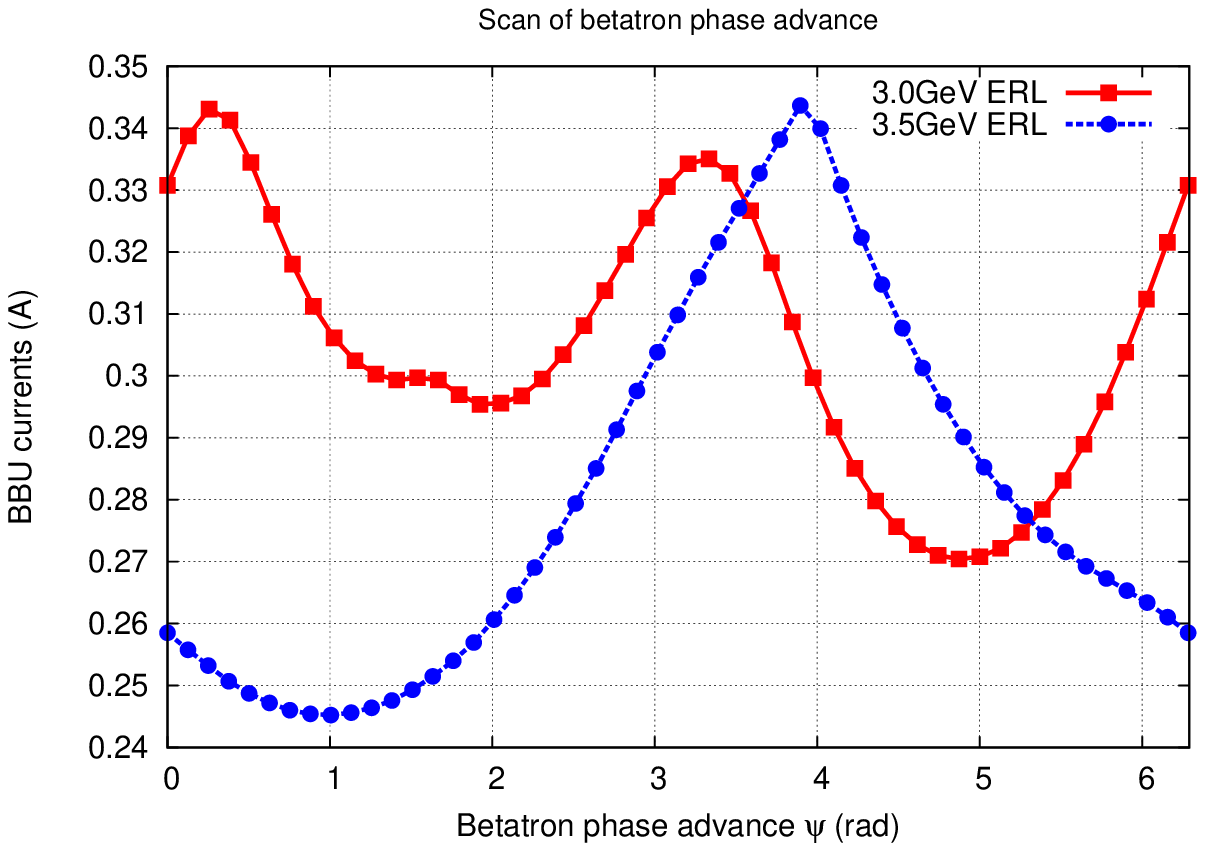}
\figcaption{\label{fig6}   BBU threshold current of two existing design of lattice. }
\end{center}

Fig.~\ref{fig6} shows the BBU threshold current as a function of the betatron phase advance for both the 3.01-GeV and the 3.41-GeV linac configurations. The HOM parameters shown in Table~\ref{tab1} are used in the calculation. The maximum BBU threshold current is found to be about 342 mA for the 3.01-GeV configuration and 343 mA for the 3.41-GeV configuration. The minimum BBU threshold current is 270 mA for the 3.01-GeV configuration and 244 mA for the 3.41-GeV configuration. The BBU threshold currents of both the two configurations meet the requirement of 100 mA average current.


\subsection{HOM randomization}
The previous simulation is based on the assumption that all cavities have the same HOM parameters. However, according to the simulation and experimental measurement\cite{lab12}, the randomization of both HOM frequency and external quality factor ($Q_{ext}$) due to cavity shape error are naturally introduced in the fabrication and tuning process of superconducting cavities. The frequency randomization reduces the coherent excitation of dipole modes in the cavity string and consequently improves the BBU threshold current. In order to simulate the influence of HOM frequency randomization, we assume a Gaussian distribution with desired frequency spread width. Due to the limited cavity number, the BBU threshold current with HOM frequency randomization shows an obvious statistical fluctuation. Therefore, we calculate the BBU threshold current for the same frequency spread for 50 times and find out the average threshold current, as well as its standard deviation. Fig.~\ref{fig9} shows the result of the simulation with the frequency spread up to 2 MHz. The results show that the BBU threshold current is significantly increased as the frequency spread increases, reaching about 950 mA (in average) when $\sigma_f=2$ MHz.


\begin{center}
\includegraphics[width=7.0cm]{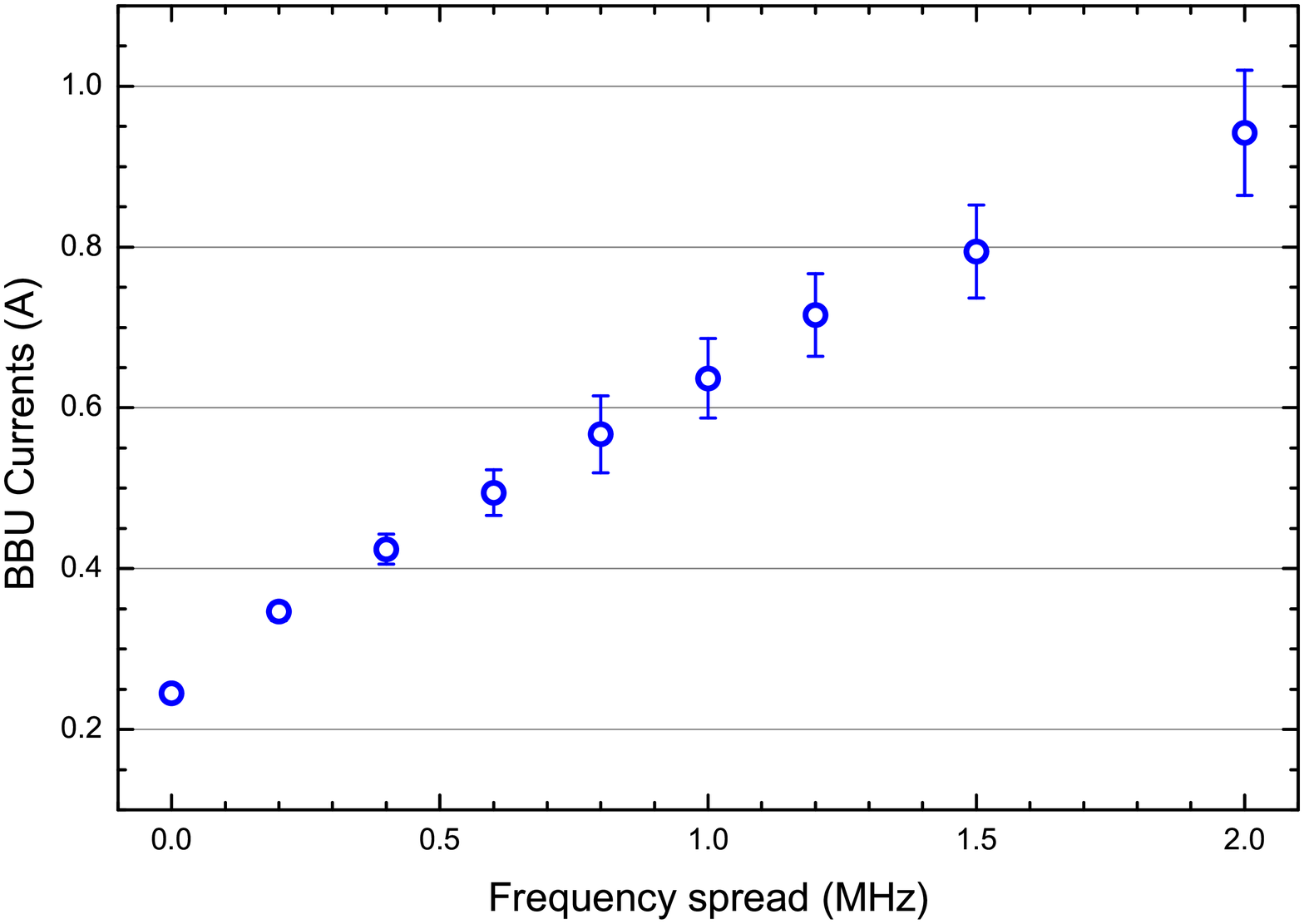}
\figcaption{\label{fig9}   Average BBU threshold current at different value of frequency spread }
\end{center}

Similar to the HOM frequency spread, the external quality factor of different cavities also shows a statistical distribution. As shown in Eq.~\ref{eq1}, the value of $Q_{ext}$ plays an essential role to the BBU instability. Therefore, the randomization of $Q_{ext}$  may impose an remarkable influence on the BBU threshold current. To investigate the influence, we assume the distribution of $Q_{ext}$ to be an uniform distribution from 0.1 to 10 times the nominal value listed in Table.~\ref{tab1}. At the same time, a Gaussian frequency distribution of $\sigma_f=2$ MHz is also applied to the HOMs so as to make the simulation close to the real situation. The BBU simulation is performed for 100 times. The statistical distribution of the BBU threshold currents for the 3.01-GeV configuration is shown in Fig.~\ref{fig10}. The result shows a broad distribution of the BBU threshold current due to the $Q_{ext}$ randomization.

\begin{center}
\includegraphics[width=6cm]{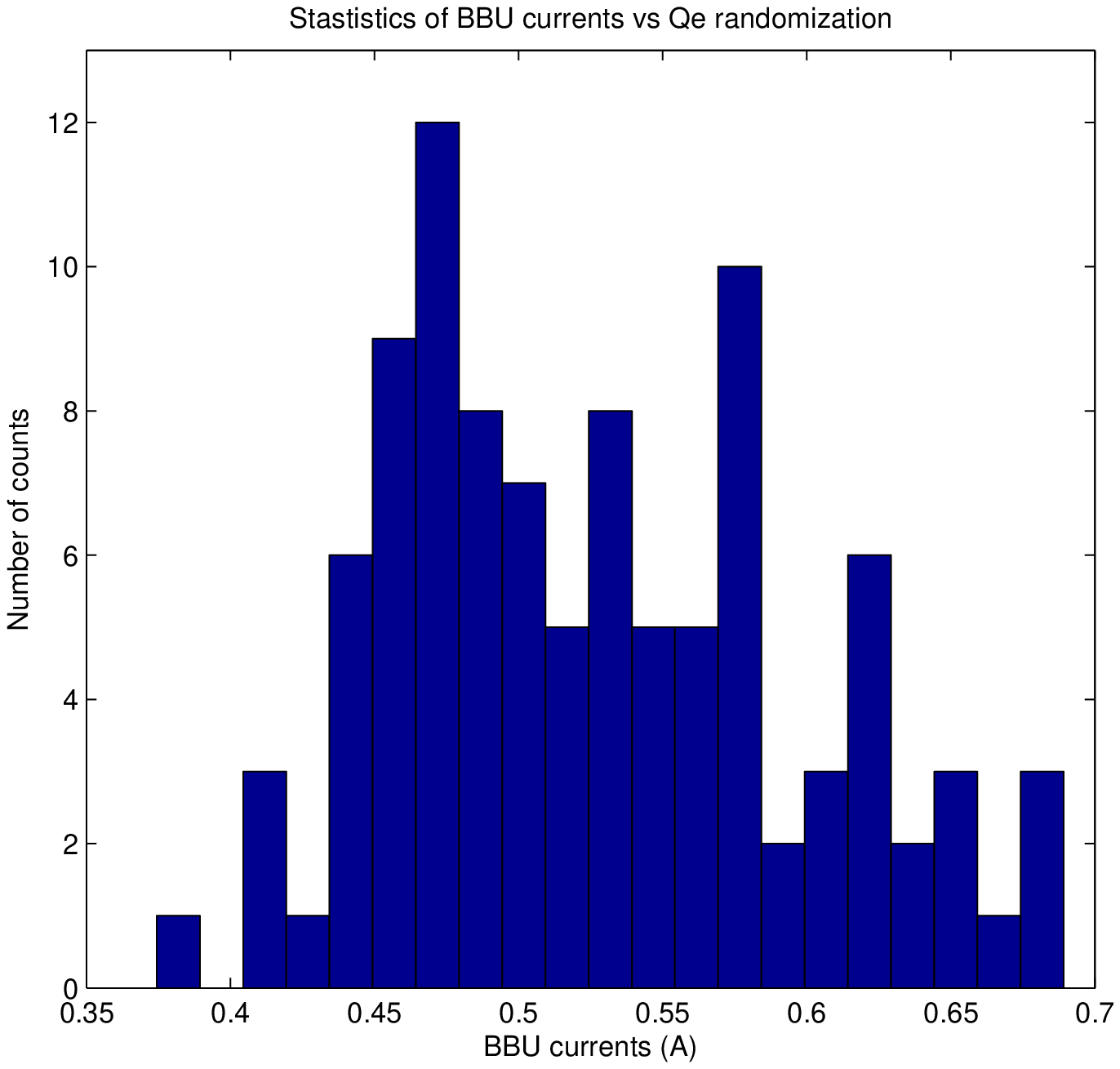}
\figcaption{\label{fig10}   Statistical distribution of the BBU threshold current with $Q_{ext}$ randomization. }
\end{center}


\subsection{Return loop length}

BBU threshold current is also a function of the recirculating loop length. The variation of $T_r$ (in Eq.~\ref{eq1}) affects the HOM phase that electron bunch experiences in the second pass through the linac. Fig.~\ref{fig11} shows the BBU threshold current versus the recirculating loop length variation, where $\Delta T/T_0$ represents the return loop length variation in terms of the relative recirculating time variation. In the case of $\sigma_f=0$, the BBU threshold current shows a periodic oscillation, which is determined by the most threatening HOM in the KEK-ERL mode-2 cavity shown in Table.~\ref{tab1}, i.e., the HOM with the frequency $f=4.011$ GHz.  In the case of $\sigma_f=1$ MHz this oscillation is smeared because the coherent excitation of this HOM is disturbed by the frequency randomization.

\begin{center}
\centering
\includegraphics[width=7.0cm]{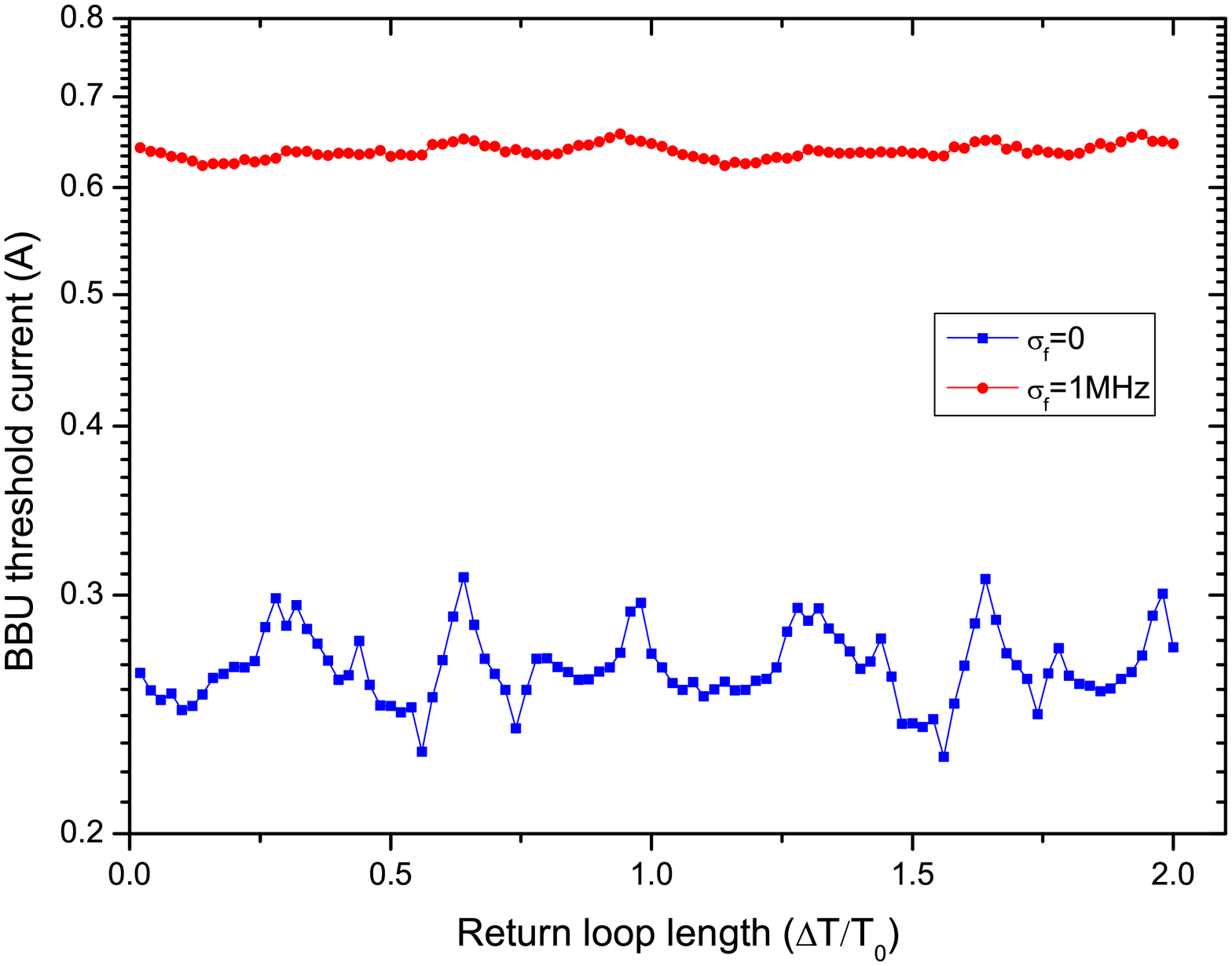}
\figcaption{\label{fig11}   BBU threshold current v.s. recirculating loop length }
\end{center}

\section{Discussion}
As discussed in Section 2, the multi-pass BBU mainly evolves from the process that a particle experience a transverse kick of the dipole HOMs when it passes through the superconducting cavity. The kick angle can be evaluated as
\begin{eqnarray}
x^\prime (y^\prime)=\frac{V_\bot}{V_p},
\label{eq3}
\end{eqnarray}
where $V_\bot$ is the transverse voltage of dipole HOM which is determined by the value of $(R/Q)Q_{ext}$, and $V_p=pc/e$ where $p$ is the beam momentum in the cavity. Thus, the HOM damping of superconducting cavity is of fundamental importance in high-energy and high-current ERLs. A previous study gives a criterion of the HOM properties to achieve 100 mA operation in an ERL \cite{lab13}
\[(R/Q)Q_{ext}/f < 1.4\times 10^5 (\Omega /cm^2/GHz),\]

As listed in Table.~\ref{tab1}, all HOMs in KEK-ERL mode-2 cavity satisfy this criterion. Therefore, a sufficient high BBU threshold current can be obtained by applying this type of cavity. In order to suppress HOMs to meet this criterion, various types of superconducting cavity have been developed all over the world, e.g., the 7-cell cavity developed for the ERL based X-ray light source at Cornell University \cite{lab14}, the 5-cell cavity developed for the ERL based e-cooling project at BNL \cite{lab15}, etc..

It can also be inferred from Eq.~\ref{eq1} and Eq.~\ref{eq3} that the cavities at low energy cavities, i.e., the cavities at the start and the end of the linac, contribute more to the BBU. We calculated the BBU threshold current of each single cryomodule in the linac of the 3.41-GeV configuration. The results are shown in Fig.~\ref{fig7}. From the figure we can see the BBU threshold currents of the first and last cryomodules are much smaller than the cryomodules in the middle of the linac. As shown in Fig.~\ref{fig6}, one can increase the BBU threshold current by adjusting the betatron phase advance of the return loop. In fact, the higher BBU threshold current occurs when the betatron phase advance makes the $T_{12}$ ($T_{34}$) value of the low energy cavities smaller. To mitigate the instability, it's also advisable to make sure the low energy cavities have smaller $Q_{ext}$ so that its contribution to the BBU can be smaller.
\begin{center}\includegraphics[width=7cm]{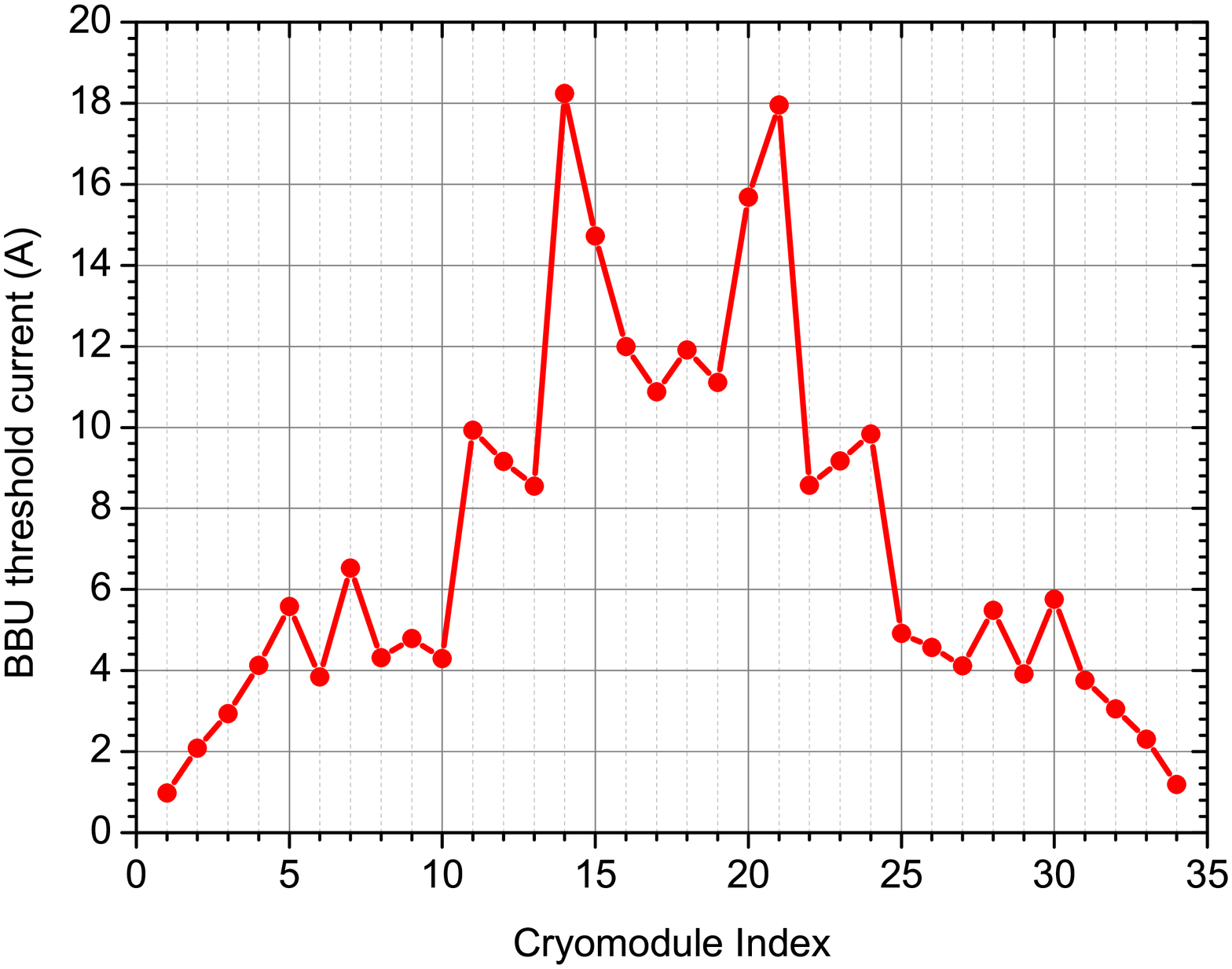}
\figcaption{\label{fig7}   BBU threshold current of each single cryomodule in the linac}\end{center}

\begin{center}
\includegraphics[width=7.0cm]{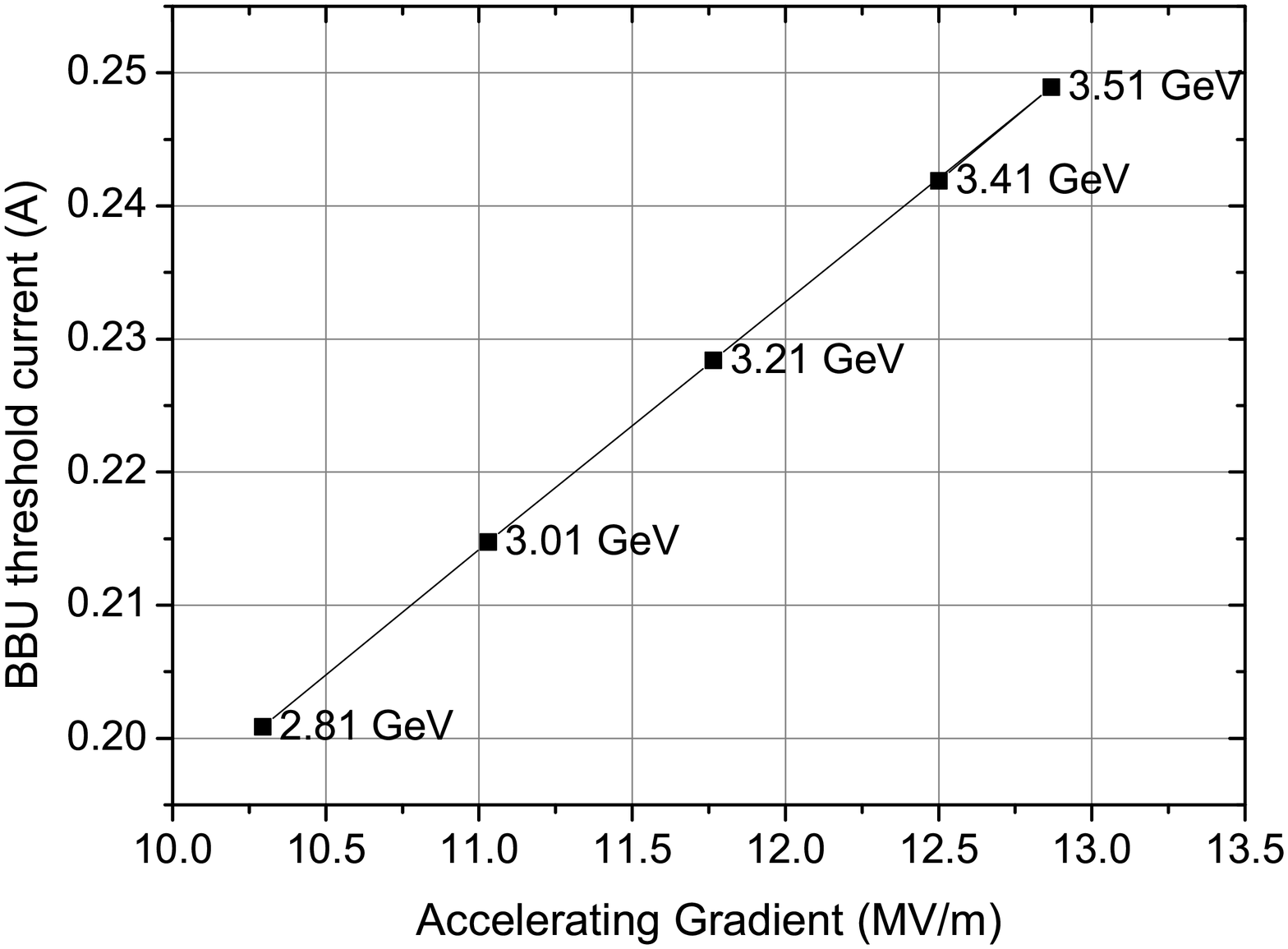}
\figcaption{\label{fig8}   BBU threshold current v.s. accelerating gradient }
\end{center}
From Eq.~\ref{eq3}, we can also infer that an obvious approach to increase the BBU threshold current is to increase the accelerating gradient of the cavity. Fig.~\ref{fig8} shows the BBU simulation of 5 ERL layouts with the same linac configuration but different accelerating gradient. A distinct increase of the BBU threshold current can be observed in the figure as the accelerating gradient increases. One can also expect a linear dependency of the BBU threshold current on the gradient of the cavity.


\section{Summary}
The transverse multi-pass BBU instability for high energy ERL has been investigated in this paper. Especially, we analyze the BBU of the KEK 3-GeV ERL light source by numerical simulation. It can be inferred from the results that the designed average current of 100 mA or more is a promising goal using the 9-cell KEK-ERL mode-2 cavity and the existing designs of linac optics. The BBU threshold current with the randomization of both HOM frequency and external quality factor are also investigated based on the simulation results. It shows that the BBU threshold current can be significantly influenced by the HOM randomization. The BBU threshold current dependance on beam energy and cavity accelerating gradient is discussed at last. The results indicate that by improving the cavity accelerating gradient, the BBU threshold current can be improved distinctly.



\section{Acknowledgment}

Most of this work is done during the visitation to KEK. Special thanks to R. Hajima, K. Ohmi, K. Umemori, Y. Yamamoto, Demin Zhou for many helpful discussions on this work. This work is also supported by the Major State Basic Research Development Program of China under Grant No. 2011CB808404.

\end{multicols}

\vspace{-1mm}
\centerline{\rule{75mm}{0.1pt}}
\vspace{2mm}

\begin{multicols}{2}

\end{multicols}

\clearpage

\end{CJK*}

\begin{thebibliography}{90}

\vspace{3mm}

\bibitem{lab1}CHEN Si, et al., Chinese Physics C, 2013()

\bibitem{lab2}CUI Xiao-Hao, et al., Chinese Physics C, 2013()

\bibitem{lab3}N. Nakamura, Proceedings of IPAC2012, p.1040-1044

\bibitem{lab4}Energy Recovery Linac Preliminary Design Report, IMSS/KEK (2012).

\bibitem{lab5}ERL Conceptual Design Report, KEK Report 2012-4

\bibitem{lab6}E. Pozdeyev, PRST-AB 8, 054401 (2005)

\bibitem{lab7}$bi$, http://www.lepp.cornell.edu/~ib38/bbu/ 

\bibitem{lab11}M. Shimada, unpublished

\bibitem{lab8}H. Sakai, et al., Proceedings of ERL2007, p.56-61

\bibitem{lab9}R. Hajima, et al., Proceedings of ERL2007, p.133-138

\bibitem{lab10}J. Rosenzweig, L. Serafini, Physical Review E, 1994, 49(2): 1599-1602


\bibitem{lab12}L. Xiao, C. Adolphsen, et al., Proceedings of PAC2007, p.2454-2456

\bibitem{lab13}M. Liepe, Proceedings of SRF2003, p.115-119
\bibitem{lab14}N. Valles, et al., Proceedings of IPAC2012, p.2384-2386
\bibitem{lab15}Wencan Xu, et al., Proceedings of PAC2011, p.2589-2591
\end{thebibliography}
\end{document}